\documentclass[a4paper,10pt]{article}
\usepackage[utf8]{inputenc}
\usepackage{authblk}
\usepackage{graphicx}
\usepackage{amssymb}
\usepackage{color}
\usepackage{amsmath}
\newcommand{\be}{\begin{equation}}
\newcommand{\ee}{\end{equation}}
\newcommand{\bea}{\begin{eqnarray}}
\newcommand{\eea}{\end{eqnarray}}
\newcommand{\nn}{\nonumber}

\newcommand{\noi}{\noindent}

\title{Soft Gluon Radiation off Heavy Quarks beyond Eikonal Approximation}

\date{}

\author[1]{Trambak Bhattacharyya \thanks{trambak.bhattacharyya@gmail.com
(Corresponding author)}}
\author[2]{Surasree Mazumder\thanks{surasree.mazumder@gmail.com}}
\author[3]{Raktim Abir\thanks{raktimabir@gmail.com}}
\affil[1]{UCT-CERN Research Centre and Department Of Physics, R W James
Building, University of Cape Town, Rondebosch 7701, Cape Town, South Africa}
\affil[2]{Instituto de Fisica, Universidade Federal do Rio de Janeiro, C.P.
68528, 21941-972, Rio de Janeiro, Brazil}
\affil[3]{Department of Physics, Aligarh Muslim University
Aligarh 202001, U.P. India}
\begin{document}

\maketitle

\begin{abstract}
We calculate the soft gluon radiation spectrum off heavy quarks (HQs)
interacting with light quarks (LQs) beyond small angle scattering
(eikonality) approximation and thus generalize the dead-cone formula of heavy
quarks extensively used in the literatures of Quark-Gluon Plasma
(QGP) phenomenology to the large scattering angle regime which may be important
in the energy loss of energetic heavy quarks in the deconfined Quark-Gluon
Plasma medium. In the proper limits, we reproduce all the relevant existing
formulae for the gluon radiation distribution off energetic quarks, heavy or
light used in the QGP phenomenology.
\end{abstract}
\section{Introduction}
High energy heavy-ion collision (HIC) programs have put Quantum Chromodynamics
(QCD), the quantum field theory of strongly interacting matters, to test in an
ambiance of high temperature ($T$) and density ($\mu$). It is of paramount
importance to measure quantities which will delineate the attributes of Quark
Gluon Plasma (QGP), a medium of deconfined quarks and gluons, expected to be
materialized in HIC in RHIC and LHC. One needs a probe to look into the
characteristics of this medium. Heavy quarks (HQs), in this context, are
believed to be very clean probes because they come to existence well before the
advent of QGP and hence they are able to watch the whole evolution of QGP.
Notwithstanding the fact that the softer part of the HQ spectrum gets
thermalized owing to its interaction with bath particles, the high frequency
counterpart sheds considerable bulk of energy which influences the experimental
observables like nuclear suppression factor ($R_{AA}$), azimuthal asymmetry
($v_2$) etc. Heavy quarks interact with thermal light quarks (LQ) /anti-quarks,
and gluons (g) mainly through elastic and/or inelastic scattering. Between the
two principal modes of energy loss, the elastic energy loss succumbs to the
radiative one in high momentum region. That is why, with increasing colliding
energies, a surge of studies in the radiative domain has been seen in past few
years \cite{GyuWang,BDMPS,Zakharov,MGMPLB,SalWiePRL,SalWie,ASW,GLV,DjoGyu,WHDG,
 Wang&Wang,WangGuo, MajumderMuller,AMY,Gossiaux,Younus,abirplb,Bass}.

 One of the main ingredients to calculate the radiative energy loss of
HQs inside QGP is the radiation spectrum. For single scattering, for example,
the radiation spectrum can be obtained by scaling the $2\rightarrow3$ inelastic
amplitude by the $2\rightarrow2$ elastic amplitude. Quantum chromodynamics based
analytical computations of radiation spectrum has so far assumed
$`$soft-eikonal-collinear' limits of parton kinematics and there is a constant
endeavour to remove the approximations. The phrase $`$soft-eikonal-collinear'
briefly conveys the following,
\\

\noi (1) {\it Soft gluons from hard partons}~:~The energy, $ E\gg\omega $ of the
emitting parton is
much larger than that of the emitted gluon, $\omega$.

\vspace{0.2cm}

\noi (2) {\it Eikonal propagation of hard jets}~:~

(a) There is no recoil of both the projectile as well as of the target parton,
{\it i.e.} $E \gg q_{\bot}$, where $q_{\bot}$ is the transverse momentum
transfer due to scattering (The Eikonal I approximation).

(b) Recoil effect on the leading parton due to emission of radiative gluon is
being neglected, {\it i.e.}
$ E \gg k_\bot $, where $k_{\bot}$ is the transverse momentum of the emitted
gluon (The Eikonal II approximation).
However, this is not an additional approximation since soft gluon emission, $
E\gg\omega $,
already encompasses it.

\vspace{0.2cm}
\noi (3) {\it Collinear emission of soft gluons}~:~

According to this assumption, gluons are predominantly emitted collinearly with
the parent parton.
{\it i.e.}, $\omega \gg k_\bot$.

\vspace{0.2cm}

\noi Hence, the `soft-eikonal-collinear' approximation assumes the following
hierarchy of different scales:
\begin{eqnarray}
E ~ \gg ~ \omega ~ \gg ~ k_\bot ~ , ~ q_\bot ~ \gg ~ m_{g,q}~ \gg ~
\Lambda_{QCD}
\label{eq00}
\end{eqnarray}
where $m_{q,g}$ is the thermal mass of quarks/gluons and $\Lambda_{QCD}\sim 200
MeV$ is the scale of QCD theory.

The radiation distribution for heavy quarks assuming some/all of the
above-mentioned approximations has been a subject matter of
Refs. \cite{dokshitzer,khardead,abirdcone}.
Ref. \cite{khardead} shows
that the HQ radiation spectrum ($dP_{\mathrm{HQ}}$) is related to LQ spectrum
($dP_{\mathrm{LQ}}$) by,
\bea
dP_{\mathrm{HQ}}=\left(1+\frac{\theta_0^2}{\theta^2}\right)^{-2}
dP_{\mathrm{LQ}},
\label{HQpower}
\eea
where $\theta$ is the radiation angle; and $\theta_0$ is the
ratio of mass of heavy quark ($m$) to its energy ($E$). However, Eq.
\ref{HQpower} assumes small angle ($\mathrm{sin}~ \theta \sim \theta$ ) limit of
the relation $k_{\bot}=\omega ~\mathrm{sin}~ \theta$, where $k_{\bot}$ and
$\omega$ are the transverse momentum and energy of bremsstrahlung gluon
respectively.
The factor
 \be
 \mathcal{D}=\left(1+\frac{\theta_0^2}{\theta^2}\right)^{-2}
 \label{dokdead}
 \ee
in Eq. \ref{HQpower} is the celebrated `dead-cone' factor.

Refs.\cite{dokshitzer,khardead} used soft-eikonal-collinear approximations
while finding out the dead-cone factor. Ref. \cite{abirdcone} has attempted to
find out the gluon radiation spectrum, and hence the heavy quark dead-cone
factor in HQ(Q)-LQ(q)$\rightarrow$HQ(Q)-LQ(q)-gluon(g) collision removing the
collinearity approximation. They have obtained a collinearity removed
`dead-cone' factor which is given by:

\be
\mathcal{D}_{NC}= \left( 1+\frac{m^2}{s} e^{2\eta} \right)^{-2}
\label{dnc}
\ee
where the subscript `NC' denotes the `non-collinear'.

It is believed that due to the presence of the `dead-cone' around the direction
of propagation heavy quark, the energy loss of heavy quarks inside medium
becomes different from those of light quarks. Since heavy quark energy loss is
related with experimental observables characterizing the QGP, a precise estimate
of the heavy-quark radiation distribution; and hence, the dead-cone factor is
essential.

Earlier the radiative energy loss of heavy quarks considering the Gunion-Bertsch
formula \cite{GB} and a modified kinematics for heavy quarks has been
calculated in \cite{MGMPLB}. The dead-cone factor in Eq. \ref{HQpower} has been
used while finding out the heavy-quark energy loss inside QGP medium
\cite{DasPRC,MBAD}. Very recently, the non-collinear soft gluon radiation
distribution containing the factor $\mathcal{D}_{NC}$ has been used while
calculating the heavy quark energy loss inside QGP in
\cite{abirplb,saraswatnpa}.

So, the latest calculation of heavy quark radiative energy loss is free from
non-collinear approximation of the emitted gluon. Though the radiative energy
loss calculations is free from the assumption of collinearity, the
Eikonal I approximation, {\it i.e.} neglecting recoil of heavy-quarks due to
scattering with medium particles, still lingers. The Eikonal I approximation
will be removed once we consider the non-negligible value of transverse momentum
transfer $q_{\bot}$ with respect to the energy $E_1$ of the incident heavy
quark. In the calculations in centre of momentum frame (COM frame), $q_{\bot}$
is related to the Mandelstam variable $t$ and the energy $E_1$ is related to
Mandelstam variable $s$. Hence, the consideration of the $\mathcal{O}(t/s)$
terms in the matrix elements calculated in Ref. \cite{abirdcone} will enable
us remove the Eikonal I approximation.

The present manuscript attempts to revisit the calculations of soft gluon (g)
radiation spectrum off heavy quarks (Q) scattering with light quarks (q)
when the recoil of heavy quark due to scattering is not negligible {\it i.e.}
when the Eikonal I approximation is not applicable any more. The hierarchy of
energy scales used is the following: \\
\\
$E\sim q_{\bot}>>\omega\sim k_{\bot}>>m_{q,g}>>\Lambda_{QCD}$. \\
\\
With the help of this calculation, \\
\\
a) We generalize the non-eikonal soft gluon radiation spectrum already
existing in Ref. \cite{abirnoneikonal} for light quarks where the effect of the
removal of Eikonal I approximation is expected to be more pronounced. \\
\\
b) We show that the eikonal formula (Eq. \ref{dnc}) in eikonal I limit of heavy
quark is reproduced.\\
\\
c) We get back the Gunion-Bertsch radiation distribution formula for
massless quarks\cite{GB}\\
\\
d) We get back he Dokshitzer-Kharzeev formula (Eq. \ref{dokdead}) in
soft-eikonal-collinear limit\\
\\
e) We provide an estimate of the effect of the large-angle scattering on
the energy loss

This manuscript is organized as follows: In the next section we describe
in detail the Feynman diagrams we use and the kinematic variables necessary for
describing our calculations. To compare with the previous works, we consider
$\mathcal{O}(g^3)$ Feynman diagrams, where $g^2=4\pi\alpha_s$ and
$\alpha_s$ is the strong coupling. The kinematic approximations will also be
discussed at length. In section \ref{radmatel} we write down the possible
Feynman amplitudes for the process in terms of the kinematic variables discussed
in section \ref{kinvar}, derive the $Qq\rightarrow Qqg$ amplitude in terms of
them and find out the non-eikonal gluon radiation spectrum. In section
\ref{plot}, we show the plots of the radiation distribution function and
show the effect of non-eikonality on radiation spectrum. In section \ref{kinreg}
we demonstrate that the present formula generalizes all the existing heavy
quark single scattering radiation distribution formulae
\cite{dokshitzer,khardead,abirdcone,GB,abirnoneikonal} used so far by taking
relevant kinematic limits. In section \ref{eloss} we calculate energy loss of
heavy/light quarks undergoing large-angle scattering while interacting with
other (light) quarks in the medium and compare with those obtained using the
results available in the literature. In the last section we summarize, draw
conclusions and attempt to mention some applications of the results obtained.

\section{Notations and Approximations:}
\label{kinvar}

\begin{figure}[ht]
\begin{center}
\includegraphics[scale=0.6]{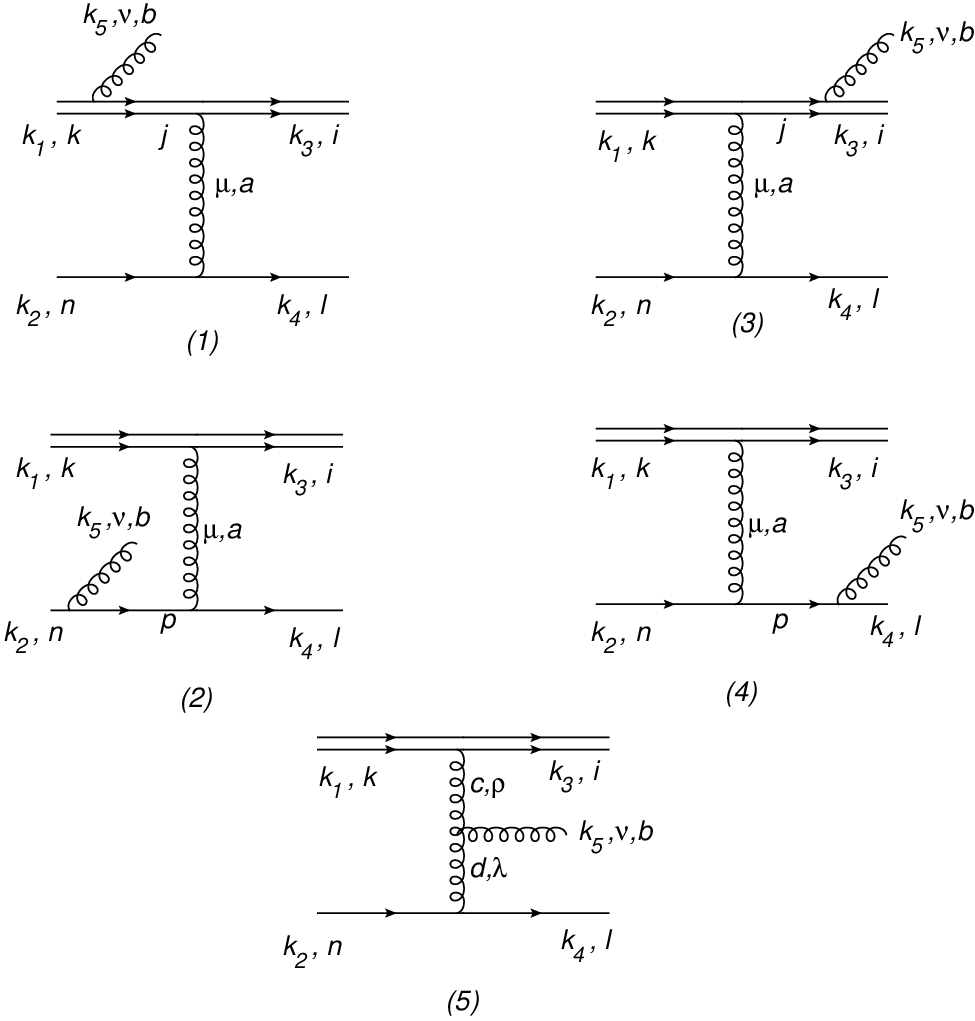}
\caption{Feynman diagrams corresponding to the process Qq$\rightarrow$Qqg.
Double line denotes heavy quarks. i, j, k, l, n, p are all quark colours. a, b,
c, d are gluon colours and Greek indices denote gluon polarizations.}
\label{diaQqQqg}
\end{center}
\end{figure}

It is well known that the gluon radiation spectrum in
Q($k_1$)q($k_2$)$\rightarrow$Q($k_3$)q($k_4$)g($k_5$)
process is given by the ratio of radiative amplitude square to the collisional
amplitude square. So our aim will be to calculate
$|\mathcal{M}_{\mathrm{Qq}\rightarrow \mathrm{Qqg}}|^2$ relaxing the eikonal
approximation due to scattering.  The Feynman diagrams contributing to the
radiative process are shown in
Fig. \ref{diaQqQqg}.

For the 2$\rightarrow$3 process obeying the four
momentum conservation relation $k_1+k_2=k_3+k_4+k_5$,
we have six Mandelstam variables $s$, $s'$, $t$, $t'$, $u$, $u'$
where
\bea
s&=&(k_1+k_2)^2,\,\,\,\, t=(k_1-k_3)^2\nonumber\\
u&=&(k_1-k_4)^2,\,\,\,\, s^{\prime}=(k_3+k_4)^2\nonumber\\
t^{\prime}&=&(k_2-k_4)^2,\,\,\,\, u^{\prime}=(k_2-k_3)^2,
\eea
subject to the constraint equation,
\be
s+t+u+s'+t'+u'=4m^2.
\ee
Hence, we need five variables
for 3-body phase space. At this point, we may assume
the four-momentum of the emitted gluon, $k_5$, to be small
enough so that the corresponding kinematics reduces to one due
to 2$\rightarrow$2 scattering. This approximation is called
the `soft gluon emission approximation'. The simplification of kinematics due
to soft gluon emission ($k_5\rightarrow0$) approximation has been discussed in
detail in \cite{DA,GR,bmad}.
In $k_5\rightarrow0$ approximation, $s\rightarrow s'$, $t\rightarrow t'$ and
$u\rightarrow u'$ which
lead to
\be
s+t+u=2m^2
\ee
Hence, the kinematics we are dealing with,
is approximately similar to the two-body kinematics which
need two Mandelstam variables, $s$ and $t$ (say), square of COM scattering
energy and COM scattering angle respectively, to be specified.
We may write down $s=2 E_1^2-m^2+2 E_1 \sqrt{E_1^2-m^2}$ and
$t=-(s-m^2)(1-\cos \theta_{13,CM})/2s$, in COM frame in terms of mass ($m$) and
energy ($E_1$) of heavy quark; and $\theta_{13,CM}$ is the COM scattering angle
between the incoming HQ (momentum $k_1$) and the scattered HQ (momentum $k_3$).
We can form, for 2-body scattering processes, two
dimensionless variables from the available quantities of our
present problem. One is $m/\sqrt{s}$ and another is $t/s$. Besides,
there may be another quantity, $k_5/\sqrt{s}$, which remind
us of the fact that we are dealing with a 3-body phase space,
in reality. Now, $k_5=(\omega,~\vec{k}_{\bot},~k_z)$; and from the previous
section we know that $|\vec{k}_{\bot}|=k_{\bot}=\omega~sin~\theta$, where
$\theta$ is
the angle the radiation makes with the parent quark. Also,
$k_z=\omega~\cos~\theta$  for on-shell radiated gluon. Consequently, all the
components of $k_5$ are now expressible in terms of $\omega$; and the third
dimensionless quantity $k_5/\sqrt{s}$ becomes proportional to $\omega/\sqrt{s}$.
Assuming $\omega/\sqrt{s}\rightarrow0$, we consider the soft limit of emitted
gluon. Under this approximation, we explore the effect of non-eikonal
contributions, {\it i.e.} $\mathcal{O}(t/s)$ terms and higher in Feynman
amplitude.


All our calculations are done in the COM frame. We hereby specify
our choice of four momenta of interacting particles. Assuming that the incoming
particles have no transverse momentum, {\it i.e.} they are travelling along the
z-axis,
say, we stick to the following choice of four momenta $k_i,~i=1\rightarrow5$.

\bea
k_1\equiv(E_1,\vec{0}_{\bot},k_{1z}),\,\,\,\,
k_2\equiv(E_2,\vec{0}_{\bot},-k_{1z}),\nn\\
k_3\equiv(E_3,\vec{q}_{\bot},k_{3z}),\,\,\,\,
k_4\equiv(E_4,-\vec{q}_{\bot},-k_{3z})\nn\\
k_5\equiv(\omega,~ \omega \sin ~\theta\hat{k}_{\bot} , ~\omega~\cos~\theta)
\label{ki}
\eea
The scattered particles are assumed to acquire a transverse momentum $q_{\bot}$.
Since we are working in COM frame in the soft gluon radiation limit, we may
approximately
assume $E_{1(2),CM}\approx E_{3(4),CM}$ and $|{\vec{p}}_{1(2),CM}|\approx
|{\vec{p}}_{3(4),CM}|$, where
approximation sign is replaced by equality for $2\rightarrow2$ case.


\section{Radiative matrix elements of HQs:}
\label{radmatel}

There are five Feynman diagrams pertaining to the process under discussion,
Qq$\rightarrow$Qqg. Obeying the standard practice (\cite{abirdcone}),
we denote a generic matrix element,
\bea
\mathcal{M}_{\alpha \beta}=\mathcal{M}_{\alpha}\mathcal{M}_{\beta}^{\dagger};
~\alpha,~\beta=1\rightarrow 5
~\forall~\alpha \le \beta
\eea
Clearly, $\alpha$ (or $\beta$) denotes the Feynman diagram being indicated among
five of them (Fig. \ref{diaQqQqg}). Below,
we list down the matrix elements, $\mathcal{M}_{\alpha \beta}$, up to terms
$\mathcal{O}(1/\omega^2)$ with
all large $t$ corrections in $\mathcal{M}$. For $\mathcal{M}_{\alpha \beta}$
with $\alpha \neq \beta$ we jot down $\mathcal{M}_{\alpha \beta}^{\mathcal{S}},
\forall \alpha \leq \beta$, where $\mathcal{M}_{\alpha \beta}^{\mathcal{S}}=
\mathcal{M}_{\alpha \beta}+\mathcal{M}_{\beta \alpha}$. $\mathcal{M}_{\alpha
\beta}=\mathcal{M}_{\beta \alpha}$, in point of fact, and hence
$\mathcal{M}_{\alpha \beta}^{\mathcal{S}}=2 \mathcal{M}_{\alpha \beta}$.

\bea
\mathcal{M}_{11}&=&\frac{128}{27}g^6
\frac{s^2}{t^2}\frac{1}{\omega^2}\frac{1}{\sin^2\theta}
\left( \frac{-1}{\tan^2\frac{\theta}{2}} \right) \mathcal{J}^2 \left( \Delta_M
^2 + \frac{f_1}{(1-\Delta_M^2)^2} \right) \nn\\
\mathcal{M}_{33}&=&\frac{128}{27}g^6
\frac{s^2}{t^2}\frac{1}{\omega^2}\frac{1}{\sin^2\theta}
\left( \frac{-1}{\tan^2\frac{\theta}{2}} \right) \mathcal{J}^2
\left( \frac{\Delta_M ^2+\frac{f_1}{(1-\Delta_M^2)^2}}
{\mathcal{F}_{35}^2}\right) \nn\\
\mathcal{M}_{13}^{\mathcal{S}}&=&\frac{128}{27}g^6
\frac{s^2}{t^2}\frac{1}{\omega^2}\frac{1}{\sin^2\theta}
\frac{1}{4} \left( \frac{-1}{\tan^2\frac{\theta}{2}} \right) \mathcal{J}^2
\left( \frac{\Delta_M ^2-\frac{f_2}{(1-\Delta_M^2)^2}} {\mathcal{F}_{35}}\right)
\nn\\
\mathcal{M}_{12}^{\mathcal{S}}&=&\frac{128}{27}g^6
\frac{s^2}{t^2}\frac{1}{\omega^2}\frac{1}{\sin^2\theta}
\frac{1}{4} (1-\Delta_M^2) \mathcal{J} \left( 1-\frac{f_3}{(1-\Delta_M^2)^3}
\right) \nn\\
\mathcal{M}_{34}^{\mathcal{S}}&=&\frac{128}{27}g^6
\frac{s^2}{t^2}\frac{1}{\omega^2}\frac{1}{\sin^2\theta}
\frac{1}{4} (1-\Delta_M^2) \mathcal{J} \left(
\frac{1-\frac{f_3}{(1-\Delta_M^2)^3}} {\mathcal{F}_{35} \mathcal{F}_{45}}\right)
\nn\\
\mathcal{M}_{14}^{\mathcal{S}}&=&\frac{128}{27}g^6
\frac{s^2}{t^2}\frac{1}{\omega^2}\frac{1}{\sin^2\theta}
\frac{7}{8} (1-\Delta_M^2) \mathcal{J} \left(
\frac{1+\frac{f_4}{(1-\Delta_M^2)^3}} {\mathcal{F}_{45}}\right) \nn\\
\mathcal{M}_{23}^{\mathcal{S}}&=&\frac{128}{27}g^6
\frac{s^2}{t^2}\frac{1}{\omega^2}\frac{1}{\sin^2\theta}
\frac{7}{8} (1-\Delta_M^2) \mathcal{J} \left(
\frac{1+\frac{f_4}{(1-\Delta_M^2)^3}} {\mathcal{F}_{35}}\right) \nn\\
\mathcal{M}_{24}^{\mathcal{S}}&=&\frac{128}{27}g^6
\frac{s^2}{t^2}\frac{1}{\omega^2}\frac{1}{\sin^2\theta}
\frac{1}{8}  \frac{t}{s} ~\tan^2~\frac{\theta}{2}~
\left( \frac{1+\frac{\frac{t}{s} \left(1+\frac{t}{2s} \right)
}{(1-\Delta_M^2)^2}} {\mathcal{F}_{45}}\right), \nn\\
\label{mij}
%
\eea
$\mathcal{M}_{22}=\mathcal{M}_{44}=0$; and $\mathcal{M}_{i5},~\forall
i=1\rightarrow 5$ do not contribute in
$\mathcal{O}(1/\omega^2)$.
The definitions of the quantities used in describing the matrix
elements in Eq. \ref{mij} are written below:

\bea
\Delta_M=\frac{m}{\sqrt{s}}~;\,\,\,
\mathcal{J}=
\frac{1-\Delta_M^2}{1+\frac{\Delta_M^2}{\tan^2~\frac{\theta}{2}}}~;\nn\\
f_1= \Delta_M^2\frac{t}{s}\left(1+\frac{t}{2s}\right)~;\,\,\,
f_2= \frac{\Delta_M^4 t}{2s}-2 \frac{\Delta_M
^2t}{s}+\frac{t}{2s}-\frac{\Delta_M ^2t^2}{2s^2}
+\frac{t^2}{2s^2}+\frac{t^3}{4s^3}  ~;\nn\\
f_3=\Delta_M^2 \frac{t}{s}-\frac{t}{s}-\frac{t^2}{2s^2}+\frac{\Delta_M^2
t^2}{2s^2}~;
\,\,\,f_4=\Delta_M^4 \frac{t}{s}-3\Delta_M^2 \frac{t}{s}+2
\frac{t}{s}-\frac{\Delta_M^2 t^2}{2s^2}+\frac{3t^2}{2s^2}
+\frac{t^3}{2s^3}~;\nn\\
\mathcal{F}_{35}= 1+\frac{  \left[ \cot~\theta \left(1 - \sqrt{1- \frac{4
\left(\frac{q_{\bot}}{\sqrt{s}}\right)^2}
{(1-\Delta_M^2)^2} }\right)
-\frac{2\left(\frac{q_{\bot}}{\sqrt{s}}\right)}{(1-\Delta_M^2)}\right]
(1-\Delta_M^2) }
{ \tan~\frac{\theta}{2} \left( 1 +
\frac{\Delta_M^2}{\tan^2~\frac{\theta}{2}}\right)     } \nn\\
\mathcal{F}_{45}= 1-\frac{  \left[ \cot~\theta
\left(1 - \sqrt{1- \frac{4\left(\frac{q_{\bot}}{\sqrt{s}}\right)^2}
{(1-\Delta_M^2)^2} }\right)
-\frac{2\left(\frac{q_{\bot}}{\sqrt{s}}\right)}{(1-\Delta_M^2)}\right]
(1-\Delta_M^2) }
{ \cot~\frac{\theta}{2} }
\label{f}
\eea
In the COM frame,
\bea
\frac{t}{s} =- \frac{q_{\bot}^2}{{s}} - \frac{1}{4}
\left(1-\Delta_M^2\right)^2
\left( 1- \sqrt {1-\frac{4
\frac{q_{\bot}^2}{s}}{\left(1-\Delta_M^2\right)^2}}\right)^2
\label{reltqperp}
\eea
Now, to define the total matrix element,
$\mathcal{M}_{\mathrm{Qq}\rightarrow\mathrm{Qqg}}$, we
need the following functions obtainable from Eq. \ref{f},

\bea
A&=&\Delta_M ^2+\frac{f_1}{(1-\Delta_M^2)^2}~;\,\,B=\Delta_M
^2-\frac{f_2}{(1-\Delta_M^2)^2}~\,\,
C=1-\frac{f_3}{(1-\Delta_M^2)^3}~;\nn\\
D&=&1+\frac{f_4}{(1-\Delta_M^2)^3}
\label{abcde}
\eea
\bea
\left|\mathcal{M}_{\mathrm{Qq}\rightarrow \mathrm{Qqg}}\right|^2 &=&
\frac{128}{27} g^6 \frac{s^2}{t^2} \frac{1}
{\omega^2\sin^2\theta} \nonumber\\
&\times& \left[ \frac{ \mathcal{C}_{1} (1-\Delta_{M}^2)^2}{ \left( 1+\frac{
\Delta_{M}^2 } {\mathrm{tan}^2 \frac{\theta}{2} }  \right) }
+ \frac{ \mathcal{C}_{2} (1-\Delta_{M}^2)^2} { \mathrm{tan}^2 \frac{\theta}{2}
\left( 1+\frac{ \Delta_{M}^2 }
{\mathrm{tan}^2 \frac{\theta}{2} } \right)^2 } + (1-\Delta_{M}^2)^2
\mathcal{C}_{0} \mathrm{tan}^2 \frac{\theta}{2}
\right]\nn\\
\label{mateleQqQqg}
\eea
where $\mathcal{C}_1$, $\mathcal{C}_2$ and $\mathcal{C}_0$ are given by,

\bea
{\cal C}_{2} &=&
-\left(A+\frac{A}{\mathcal{F}_{35}^2}+\frac{B}{4\mathcal{F}_{35}}\right)~;  \nn
\\
{\cal C}_{1} &=&\frac{C}{4}
\left(1+\frac{1}{\mathcal{F}_{35}\mathcal{F}_{45}}\right)
+\frac{7 }{8}D \left(\frac{1}{\mathcal{F}_{45}}+\frac{1}{
\mathcal{F}_{35}}\right)~; \nn \\
{\cal C}_{0} &=&\frac{1}{8\mathcal{F}_{45}(1-\Delta_M^2)^4}
\left[(1-\Delta_M^2)^2\frac{t}{s} + \frac{t^2}{s^2} +\frac{1}{2} \frac{t^3}{s^3}
\right]~; \nn \\
\label{C_n}
\eea
Using gluon rapidity $\eta=-\text{ln}\left(\text{tan}\frac{\theta}{2}\right)$
and the light cone variable
$x=k_{\bot}e^{\eta}/\sqrt{s}$, we can get

\bea
\left| \mathcal{M}_{\mathrm{Qq}\rightarrow \mathrm{Qqg}} \right |^2=
\frac{16}{3} g^2 \left| \mathcal{M}_{\mathrm{Qq}\rightarrow \mathrm{Qq}} \right
|^2 \frac{1}{\omega^2}\frac{1}{\sin^2\theta}
\underbrace{\left[\sum_{n=2,1,0}^{}{\cal C}_{n}~e^{2(n-1)\eta
}\left(\frac{k_\bot^{2}}
{k_\bot^2+x^2M^2}\right)^{n} \right]}_{W(x,k_{\bot}^2)}\nn\\
\label{mateleQqQqgnew}
\eea
Where we use,
\bea
\left| \mathcal{M}_{\mathrm{Qq}\rightarrow \mathrm{Qq}} \right |^2=
\frac{8}{9}g^4\frac{s^2}{t^2} (1-\Delta_M^2)^2
\eea
$W$ is related with the radiation spectrum off HQs when the Eikonal I
approximation is removed.

\section{The non-eikonal radiation spectrum off Heavy Quarks}
\label{plot}

\begin{figure}[h]
\begin{center}
\includegraphics[scale=0.6]{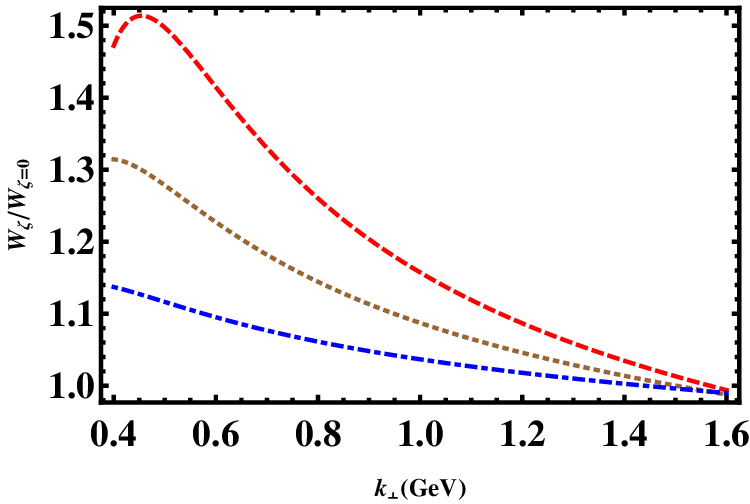}
\caption{Variation of the non-eikonal radiation spectrum scaled by the eikonal
spectrum ($\zeta$=0.0) off heavy quark ($\Delta_M=0.1$) with gluon transverse
momentum. Red(dashed): $\zeta$=0.15; Brown(dotted): $\zeta$=0.30;
Blue(dot-dashed): $\zeta$=0.45}
\label{spectrumratioplot}
\end{center}
\end{figure}

In Fig. \ref{spectrumratioplot} we show the variations of the non-eikonal
spectra ($W_{\zeta}$) scaled by the eikonal spectrum ($W_{\zeta=0}$) with
respect to the gluon transverse momentum $k_{\bot}$. We see that for soft
approximation and for comparatively less non-eikonality ($\zeta=0.15$) the
contribution due to non-eikonality may be 50\% more than that due to
eikonality. This excess may reach up to $\sim$30 \% ($\sim$15 \%)for
$\zeta=0.30$($\zeta=0.45$).

\begin{figure}[h]
\begin{center}
\includegraphics[scale=0.6]{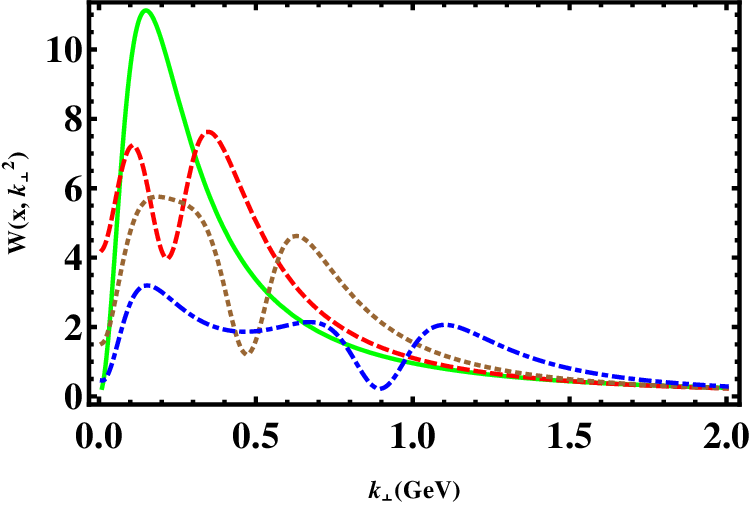}
\caption{Variation of gluon spectrum $W(x,k_{\bot}^2)$ off heavy quark
($\Delta_M=0.1,~x=0.1$) with gluon transverse momentum for different extents of
recoil of heavy quarks. Green(solid): $\zeta$=0.0; Red(dashed): $\zeta$=0.15;
Brown(dotted): $\zeta$=0.30; Blue(dot-dashed): $\zeta$=0.45}
\label{spectrumplot}
\end{center}
\end{figure}

In Fig. \ref{spectrumplot} we plot the non-eikonal radiation
spectrum off heavy quarks, $W$, with varying $k_{\bot}$ of gluons for different
$\zeta$ values. $\zeta=q_{\bot}/\sqrt{s}$ signifies the extent of transverse
momentum transferred to the heavy quark due to scattering with light quarks.
Hence, $\zeta$ can be treated as the non-eikonality parameter in our
calculation.

If we want to calculate the energy loss and its effect on the nuclear
suppression factor we have to consider the $Qg\rightarrow Qgg$ scattering which
will have more cross-section than the $Qq\rightarrow Qqg$, too. While in the
eikonal case \cite{abirdcone}, the $Qg\rightarrow Qgg$ matrix element differs
from that of $Qq\rightarrow Qqg$ just by a number due to color factor, the
non-eikonal case is not going to be so simple and we have to calculate the
Feynman amplitudes of a lot more diagrams. The present calculation may, in
principle, be useful when quarks dominate in the medium. But that needs a
consistent treatment of the multiple scattering process. Once that is done, we
can easily find out the effect of non-eikonality in energy loss.

\section{Behaviour of non-eikonal heavy quark spectrum at different kinematic
regions:}
\label{kinreg}

\subsubsection*{Region I: Massless quark with non-eikonal trajectory}

In the massless limit of Eq. \ref{mateleQqQqg}, we obtain the non-eikonal gluon
radiation spectrum off light quarks.
Below we jot down the forms of the functions $f_i,\forall i=1\rightarrow5$,
$A\rightarrow D$ and $\mathcal{C}_1,
\mathcal{C}_2,\mathcal{C}_0$ when we take massless limit, {\it i.e.}
$m\rightarrow0\Rightarrow\Delta_M\rightarrow0$,

\bea
&&(\text{i}) \mathcal{J}\rightarrow1 \nn\\
&&(\text{ii})~f_1\rightarrow0~;\,f_2\rightarrow\frac{t}{2s}+\frac{t^2}{2s^2}
+\frac{t^3}{4s^3}~;\nn\\
&&\,\,f_3\rightarrow-\frac{t}{s}-\frac{t^2}{2s^2}~;\,\,f_4\rightarrow\frac{2t}{s
}+
\frac{3t^2}{2s^2}+\frac{t^3}{2s^3}\nn\\
(\text{iii}) ~\mathcal{F}_{35}\rightarrow\mathcal{F}^0_{35}&=&
1+\left[ \cot\theta \left(1-\sqrt{1-4\frac{q_{\bot}^2}{s}}\right) -
\frac{2q_{\bot}}{\sqrt{s}} \right] \cot\frac{\theta}{2}~;\,\,\nn\\
\mathcal{F}_{45}\rightarrow\mathcal{F}^0_{45}&=&
1+\left[ \cot\theta \left(1-\sqrt{1-4\frac{q_{\bot}^2}{s}}\right) -
\frac{2q_{\bot}}{\sqrt{s}} \right] \tan\frac{\theta}{2}\nn\\
&&(\text{iv}) A\rightarrow0~;\,\, B\rightarrow
B^0=-\frac{t}{2s}-\frac{t^2}{2s^2}-\frac{t^3}{4s^3}~;\nn\\
&&\,\,C \rightarrow C^0=1+\frac{t}{s}+\frac{t^2}{2s^2}~;\,\,
D\rightarrow D^0=1+\frac{2t}{s}+\frac{3t^2}{2s^2}+\frac{t^3}{2s^3}\nn\\
(\text{v}) \mathcal{C}_1 &\rightarrow& \mathcal{C}_1^0=\frac{C^0}{4}+
\frac{C^0}{4\mathcal{F}^0_{35}\mathcal{F}^0_{45}}+\frac{7D^0}{8\mathcal{F}^0_{35
}}+\frac{7D^0}{8\mathcal{F}^0_{45}}~;\,\,\nn\\
\mathcal{C}_2 &\rightarrow&
\mathcal{C}_2^0=-\frac{B^0}{4\mathcal{F}^0_{35}}\nn\\
\mathcal{C}_0 &\rightarrow& \mathcal{C}_0^0=\frac{1}{8\mathcal{F}^0_{45}}
\frac{t}{s}
\left(1+ \frac{t}{s}\left(1+\frac{t}{2s}\right) \right)
\eea

Hence,
\bea
\left|\mathcal{M}_{\mathrm{qq'}\rightarrow \mathrm{qq'g}}\right|^2=12g^2
\frac{1}{k_{\bot}^2}
|M_{\mathrm{qq'}\rightarrow \mathrm{qq'}}|^2
\left \{ \mathcal{C}_1^0+ \frac{\mathcal{C}_2^0} { \tan^2 \frac{\theta}{2}} +
\mathcal{C}_0^0 \tan^2\frac{\theta}{2} \right\}
\eea

If we retain the terms up to $\mathcal{O}(\frac{t}{s})$ of $B^0,C^0,D^0$ and
put $\mathcal{F}_{35}=1=\mathcal{F}_{45}$, we get
\bea
\left|\mathcal{M}_{\mathrm{qq'}\rightarrow \mathrm{qq'g}}\right|^2 &=&
\frac{128}{27}g^6 \frac{s^2}{t^2}\frac{1}{k_{\bot}^2}
\left \{
2\frac{1}{4}\left(1+\frac{t}{s}\right)+2\frac{7}{8}\left(1+\frac{2t}{s}\right)
+ \frac{t}{8s} \frac{1}{ \tan^2 \frac{\theta}{2}} + \frac{t}{8s}
\tan^2\frac{\theta}{2} \right\} \nn\\
&=&12 g^2  \left\{ \frac{8}{9} g^4 \frac{s^2}{t^2} \right\}
\frac{1}{k_{\bot}^2}
\left(1+\frac{16t}{9s}+\frac{t}{9s}\cosh 2\eta\right)\nn\\
\label{mateleqqqqgabir}
\eea
In the limit $\eta\rightarrow 0$ Eq. \ref{mateleqqqqgabir} boils down to the
light quark non-eikonal
(up to $\mathcal{O}(t/s)$) matrix element obtained in Ref.
\cite{abirnoneikonal}.

\subsubsection*{Region II: Massive quark with eikonal trajectory}

This region considers
\bea
\frac{q_{\bot}}{\sqrt{s}}\rightarrow0 \Rightarrow \frac{t}{s}\rightarrow 0 \nn\\
\eea
Hence, from Eq. \ref{f},
\bea
f_i=0 ~\forall ~i=1\rightarrow5 ~;\,\,\mathcal{F}_{35}=\mathcal{F}_{45}=1
\label{ft0}
\eea
From Eq. \ref{abcde} we get, in the same limit,
\bea
A=B=\Delta_M^2~;\,\,C=D=1~;\,\,\mathcal{F}_{35}=\mathcal{F}_{45}=1
\label{abcdet0}
\eea
Hence
\bea
\mathcal{C}_1=\frac{9}{4}~;\,\,\mathcal{C}_2=-\frac{9\Delta_M^2}{4}~;\mathcal{C}
_0=0
\label{C_nt0}
\eea
In Eq. \ref{C_n}.

With the help of Eqs. \ref{ft0}, \ref{abcdet0}, \ref{C_nt0}
we get,

\bea
\left|\mathcal{M}_{\mathrm{Qq}\rightarrow \mathrm{Qqg}}\right|^2&=&
\frac{128}{27} g^6 \frac{s^2}{t^2} \frac{1}
{k_{\bot}^2} \frac{9}{4}\mathcal{J}^2 \nn\\
&=&\frac{128}{27} g^6 \frac{s^2}{t^2} \frac{1}
{k_{\bot}^2} (1-\Delta_M^2)^2 \frac{9}{4} \frac{1}{\left(
1+\frac{\Delta_M^2}{\tan^2~\frac{\theta}{2}} \right)^2}\nn\\
&=& 12 g^2
\left[ \frac{8}{9} g^4 \frac{s^2}{t^2} (1-\Delta_M^2)^2 \right]
\frac{1}{k_{\bot}^2} \frac{1}{\left(
1+\frac{\Delta_M^2}{\tan^2~\frac{\theta}{2}} \right)^2} \nn\\
&=& 12 g^2 \left|\mathcal{M}_{\mathrm{Qq}\rightarrow \mathrm{Qq}}\right|^2
\left\{ \frac{1}{k_{\bot}^2}
\left( 1+\frac{m^2}{s} e^{2\eta} \right)^{-2} \right\}
\label{mateleabiretal}
\eea
with $\eta=-\ln\left(\tan~\frac{\theta}{2}\right)$; and the expression embraced
by the curly
braces is the radiated gluon spectrum
($\sim {\left|\mathcal{M}_{\mathrm{Qq}\rightarrow \mathrm{Qqg}}\right|^2}/
{\left|\mathcal{M}_{\mathrm{Qq}\rightarrow \mathrm{Qq}}\right|^2}$) for this
case.
Evidently, the present calculation yields the calculation in Ref.
\cite{abirdcone} in the small
angle scattering limit (Eq. \ref{mateleabiretal}).

\subsubsection*{Region III: Massless quark with eikonal trajectory}

Now we explore the behaviour of the radiation spectrum in the following limits,
\bea
(\text{i})&&\frac{q_{\bot}}{\sqrt{s}}\rightarrow 0 \Rightarrow
\frac{t}{s}\rightarrow 0\nn\\
(\text{ii})&&m=0\Rightarrow \Delta_M=0\Rightarrow\mathcal{J}\rightarrow 1 \nn\\
\eea
The above limits force Eq. \ref{mateleabiretal} to take the form given below,
\bea
\left|\mathcal{M}_{\mathrm{qq'}\rightarrow \mathrm{qq'g}}\right|^2&=&
12 g^2 \left|\mathcal{M}_{\mathrm{Qq}\rightarrow \mathrm{Qq}}\right|^2
\frac{1}{k_{\bot}^2}
\eea
which in the limit $q_{\bot}>>k_{\bot}$ can be written as,

\bea
\left|\mathcal{M}_{\mathrm{qq'}\rightarrow \mathrm{qq'g}}\right|^2 \approx
12 g^2 \left|\mathcal{M}_{\mathrm{qq'}\rightarrow \mathrm{qq'}}\right|^2
\left[ \frac{q_{\bot}^2}{k_{\bot}^2(\vec{k}_{\bot}-\vec{q}_{\bot})^2} \right],
\eea
where $q,q'$ are two different light quark flavors. The part within the square
braces
can very well be identified with the celebrated Gunion-Bertsch gluon spectrum
\cite{GB}
emitted from light quarks.

\subsubsection*{Region IV: Massive quark with eikonal trajectory emitting
collinear gluons}
This region considers the following limits

\bea
&&\text{(i)}~~m<<\sqrt{s}\Rightarrow s\approx 4 E_1^2 \nn\\
&&\text{(ii)}~~\frac{q_{\bot}}{\sqrt{s}}\rightarrow 0 ~~\text{and} \nn\\
&&\text{(iii)}~~\theta\rightarrow 0 \Rightarrow \tan~\frac{\theta}{2}\approx
\frac{\theta}{2}
\eea
In the above limit Eq. \ref{mateleQqQqg} yields the dead-cone factor of Ref.
\cite{khardead},

\bea
\left|\mathcal{M}_{\mathrm{Qq}\rightarrow \mathrm{Qqg}}\right|^2&=&
12 g^2 \left|\mathcal{M}_{\mathrm{Qq}\rightarrow \mathrm{Qq}}\right|^2
\frac{1}{k_{\bot}^2}
\left(1+\frac{\theta_0^2}{\theta^2}\right)^{-2},
\eea
with $\theta_0=\frac{m}{E_1}$.

\section{Estimation of energy loss}
\label{eloss}

In this section we calculate the eikonal and non-eikonal energy  loss per unit
length ($\frac{dE}{dx}$  in  a medium of infinite extent) experienced by the
heavy/light quarks to estimate the quantitative difference among various
existing formulae in \cite{abirdcone,abirnoneikonal}. Here we outline the
scheme of our energy loss calculations in brief. The detailed procedures and
of the energy loss calculations can be obtained in \cite{bmad,wanggyulassy}.

We consider a thermal bath of light quarks at temperature $T=300$ MeV with which
the heavy quarks interact. The interaction of the heavy quark with the light
quarks is encoded in the Feynman amplitude calculated. Also, due to the
presence of thermal bath, the light quarks and the radiated gluons will acquire
thermal masses. We can take the quark thermal mass as $m_f^2=\pi
\alpha_s(T)T^2C_F/2$ ; and the gluon thermal mass ($m_g$) is given by
$m_g^2=2\pi \alpha_s(T)T^2 (C_A+N_f/2)/3$ \cite{bellac}. $C_A$($C_F$) is the
Casimir factor in the Adjoint(Fundamental) representation and $N_f$ is the
number of
flavours.


Energy loss (per collision) due to radiated gluons can be obtained if we
integrate the gluon spectrum, which is related to the ratio of the
2$\rightarrow$3 amplitude square to the 2$\rightarrow$2 amplitude square and is
weighted by the gluon energy ($\omega$), over the gluon transverse momentum
($k_{\bot}$) and its rapidity ($\eta$). If we restrict ourselves within the
Bethe-Heitler additive region, there will be an upper limit imposed on the
$k_{\bot}$ value. The average energy loss per unit length can be obtained if we
multiply the energy loss per collision with the collision rate, which we have
using the techniques detailed in \cite{thomaintrate}.

\begin{figure}[h]
\centering
\includegraphics[width=0.8\columnwidth, height = 8.0cm]{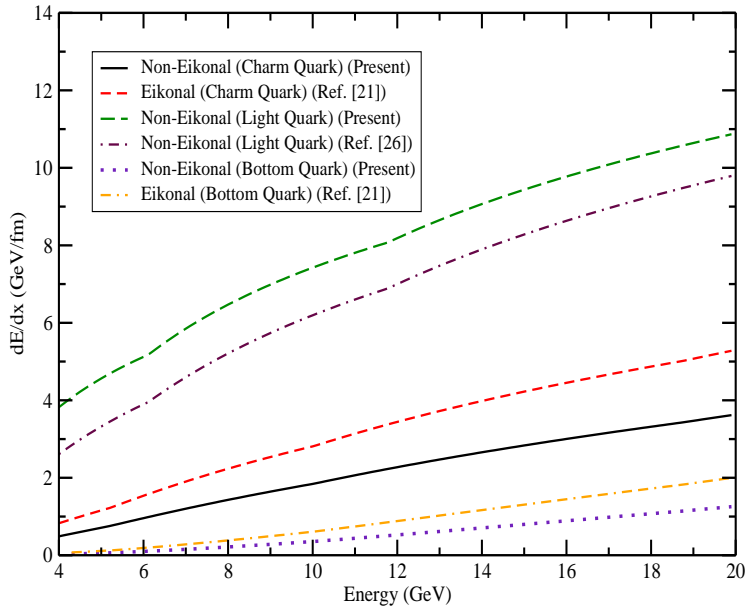}
\caption{Energy loss of quarks in a thermal bath of 300 MeV}
\label{vsE}
\end{figure}

We observe from Fig. \ref{vsE} that the inclusion of the effect of
non-eikonality can result in $\sim$ 55 \% ($\sim$ 39 \%) change in energy loss
for a 8 GeV charm quark (bottom quark) and $\sim$ 48 \%  ($\sim$ 43 \%) change
in energy loss for a 16 GeV charm quark (bottom quark).

For light quarks, the non-eikonal energy loss contains
contributions from the terms of order $t^2/s^2$ and $t^3/s^3$ in the matrix
element which are absent in the calculations of Ref. \cite{abirnoneikonal}.
So, the non-eikonal energy losses of light quarks of 8 and 16 GeV differ by 24
\% and 13 \% respectively.

\vskip 0.6in

\section{Summary and conclusion}
\label{sumcon}

In Summary, we have found out the non-eikonal radiation distribution off heavy
quarks scattering with light quarks. Also, from Fig. \ref{spectrumratioplot}, we
realize that for soft approximations we can hardly rule out the importance of
the non-eikonality. Fig. \ref{vsE} shows that the effect of non-eikonality may
be substantial for highly energetic heavy/light quarks. And, the consideration
of the effects of non-eikonality will substantially modify the phenomenology
related to the heavy-quark dynamics.

The non-eikonal distribution boils down to the all existing radiation
distribution formulae provided we choose proper kinematic limits. This analysis
will help towards the advancement of the continuous endeavour of relaxing the
kinematic limits lingering inside the calculations of energy loss.

Unlike the eikonal case, the matrix element for the Qg$\rightarrow$Qgg
process cannot be found out just by changing the color factor. The matrix
element has to be evaluated for finding out non-eikonal energy loss in RHIC and
LHC energy domains. The multiple scattering may be included inside the present
analysis taking into account the interference effects of the scattering
amplitudes due to successive collisions inside the medium. Also, recently in
Ref. \cite{birov2} the radiation pattern is shown to give rise to an azimuthal
asymmetry which does not have any hydrodynamical origin. The present
calculations may be employed to calculate the azimuthal asymmetry generated due
to non-eikonality. The observed results can be compared/contrasted with the
experimental findings; and that study will be the subject matter of an upcoming
research paper.

\vskip 0.3in
\section*{Acknowledgements}
The authors acknowledge the support of VECC and SINP, Kolkata, India where the
substantial part of the work has been done. TB acknowledges UCT-URC for support.
RA acknowledges discussion with Santanu Maity.

\end{document}